# THEORETICAL AND METHODOLOGICAL APPROACHES TO THE STUDY OF THE PROBLEM OF CORRUPTION

**Valeri Lipunov**,
Associate Professor, Law Institute of the Russian University of Transport (Russia)
**Vladislav Shirshikov,**
Associate Professor, Law Institute of the Russian University of Transport (Russia)
**Jonathan Lewis,**
Associate Professor, Hitotsubashi University (Japan)

**Abstract.**

The article reveals the main theoretical approaches to the analysis and study of the phenomenon of corruption. Special attention is paid to the consideration of the index approach to the analysis of corruption.

**Introduction**

In most states, especially in developing countries, there is concern about the development and promotion of effective anti-corruption strategies. The possibility of implementing such strategies stems from the direct link, demonstrated theoretically and empirically, between the consequences of anti-corruption strategies and the results of government activities, expressed both in economic and social results, as well as in living standards, well-being, etc.

**Methodology**

The identification of various signs of the phenomenon of corruption and the study of the process of using the corruption component in self-organizing systems has predetermined the use of the methodology of social synergy presented in the works of V. Bransky, M. Elchaninov, T. Kolesnikova, V. Romanov, G. Haken and other researchers [3; 5; 16; 22; 9].

The structural-functional approach, which is represented by the theory of social anomie of R. Merton [19], and the institutional approach, reproduced by the concept of modernization of S. Huntington [12], have been implemented in the key of determining corruption in the social system, and also considered the features of the relationship between the state of the social structure of society and the level of development of corruption relations.

The theory of rational and efficient bureaucracy by M. Weber [24] and the concept of political modernization, civic culture and stability of democracy by G. Almond [1] served as the basis for the analysis and further characterization of the relationship between the efficiency of the organization of the state apparatus, as well as the peculiarities of the relationship between citizens and officials and the level of corruption.

The works of B. Hofsted, S. Lipset and G. Lenz helped to reveal the significance of the role of cultural factors in the processes of using corruption practices [15; 18].

**Results and discussion**

In the field of sociological research of the phenomenon of corruption, there are a number of modern theoretical and methodological approaches. For sociological analysis, the definition of social subjects of corruption plays an important role. Corruption crimes can be considered as a form of social relations, since corruption usually involves several actors. The analysis of the corruption component as a form of social relations is reflected in the works of E. Giddens and Yu. Habermas [7].

Such foreign researchers as T. Parsons and R. Merton adhere to a systematic and structural-functional approach, considering the phenomenon of corruption as a systemic phenomenon that has its own structure, signs and properties [19].

To look at corruption through the prism of existing forms of interactive interaction allows the interactionist approach, represented by the works of Tolkien, Meade and Bloomer [13]. Symbolic interactionism offers a wide range of ideas based on the philosophy of pragmatism and psychological behaviorism. The latter is very valuable in the analysis of institutional manifestations of corruption.

From the point of view of the theory of exchange, which is represented by Tolkien and Homans [11], corruption is a special form of social exchange, represented as a material and symbolic exchange of resources and relationships in the community.

G. Becker and G. Stigler [2] present their vision of the problem of corruption, studying it from the point of view of the theory of rational choice, according to which in modern Russian society it shows that the adoption of socially disapproved patterns of behavior by participants in corruption relations is carried out in most cases without pressure on the individual from authoritative personalities.

The corruption phenomenon can be described in the context of network theory, which allows us to consider the phenomenon of corruption in the context of corrupt social networks formed by the subjects of corruption, both in the local community and in the hierarchy of vertically organized social management systems. In this regard, the works of L. Freeman are of special interest [6].

A slightly different theoretical approach connects various forms of corruption with the perception of the level of corruption of the elite and the public. One of the earliest studies of A. Heydenheimer [10] relates the perception of corruption to the established nature of obligations in society. In a similar approach, using data on public opinion on the issue under consideration, J. Peters and S. Welch [21] believes that actions involving a non-political official, such as a judge rather than a politician, or a public figure rather than a private citizen, donations from voters rather than constituents, and large payments are generally considered more serious.

The study by P. Lascoumes and O. Tomescu-Hatto also assesses the level of corruption in accordance with the types of corruption actions [17]. This makes it possible to distinguish certain groups of citizens based on their ideas about the degree of tolerance to certain manifestations of favoritism and the general perception of the degree of corruption among civil servants.

A significant part of the difficulties in studying various forms of corruption is related to the methodology. Despite the numerous typologies developed by analysts, there are still no clear tools for a comprehensive assessment of the various types of corruption. To date, the Transparency International Corruption Perception Index is the most frequent tool used in empirical work on the causes or consequences of corruption [4]. This indicator, however, is one-dimensional and does not differentiate the forms of corruption. Although the use of one-sided corruption indicators is important for the development of anti-corruption research, they cannot really address issues related to various types of corruption (Tab.1).

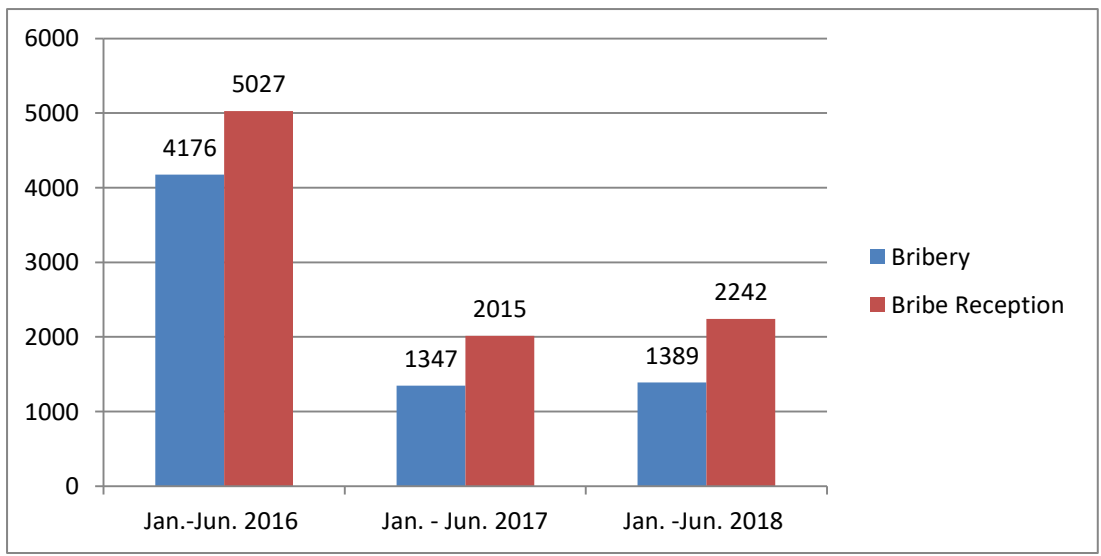

Table 1. Changes in the number of corruption crimes in Russia from 2016 to 2018 (calculated by the authors)

It should be noted that the Transparency International Corruption Perception Index is based on the perception of public opinion through opinion polls, which cannot differentiate the state of corruption on the ground. In any case, this barometer is an important methodological tool for differentiating countries not only by the overall level of corruption, but also by its various models. In this case, it falls within the scope of the theoretical task to use this data to test and refine theories that determine the cause and effect of acts of corruption.

**Conclusion**

As a conclusion, it can be noted that corruption, being a multi-faceted phenomenon, is studied by many sciences and is considered in various aspects. Practice shows that close attention is paid to the methods and forms of fighting corruption. In addition, corruption is mainly considered as a socio-political and economic phenomenon, which determines the theoretical and methodological approaches to its study.

Of course, corruption is more common where there are administrative, economic, financial and other resources, as well as the possibility of abuse of them, or difficult to solve social problems in countries experiencing a period of crisis or modernization. In this case, corruption becomes a tool for influencing the adoption and execution of public or private decisions through pressure. In this regard, anti-corruption activities occupy the attention of an increasing part of the state apparatus and public opinion, which should be addressed in the next chapter of this work.